\documentclass[amsmath,twocolumn, prl,aps]{revtex4-1} 
\usepackage{amsthm,amsfonts,graphicx,verbatim, color}
\usepackage{bm}
\usepackage[utf8]{inputenc}
\let\oldmarginpar\marginpar
\renewcommand\marginpar[1]{\-\oldmarginpar[\raggedleft\footnotesize #1]%
{\raggedright\footnotesize #1}}

\newcommand{\be}{\begin{equation}}
\newcommand{\ee}{\end{equation}}
\newcommand{\bea}{\begin{eqnarray}}
\newcommand{\eea}{\end{eqnarray}}

\renewcommand{\epsilon}{\varepsilon}
\renewcommand{\vec}[1]{{\bf #1}}

\def\beq{\begin{equation}}
\def\eeq{\end{equation}}
\def\bea{\begin{eqnarray}}
\def\eea{\end{eqnarray}}

\begin{document}

\title{Floquet Thermalization: Symmetries and Random Matrix Ensembles}
\author{Nicolas Regnault$^{1,2}$}
\author{Rahul Nandkishore$^{3,4}$}
\affiliation{$^1$ Department of Physics, Princeton University, Princeton NJ 08544, USA\\
$^2$ Laboratoire Pierre Aigrain, Ecole Normale Sup\'erieure-PSL Research
University, CNRS, Universit\'e Pierre et Marie Curie-Sorbonne Universit\'es,
Universit\'e Paris Diderot-Sorbonne Paris Cit\'e, 24 rue Lhomond, 75231
Paris Cedex 05, France\\
$^3$Department of Physics, University of Colorado, Boulder, Colorado 80309, USA\\
$^4$Center for Theory of Quantum Matter, University of Colorado, Boulder, Colorado 80309, USA}
\begin{abstract}
We investigate the role of symmetries in determining the random matrix class describing quantum thermalization in a periodically driven many body quantum system. Using a combination of analytical arguments and numerical exact diagonalization, we establish that a periodically driven `Floquet' system can be in a different random matrix class to the instantaneous Hamiltonian. A periodically driven system can thermalize even when the instantaneous Hamiltonian is integrable. A Floquet system that thermalizes in general can display integrable behavior at commensurate driving frequencies. When the instantaneous Hamiltonian and Floquet operator both thermalize, the Floquet problem can be in the unitary class while the instantaneous Hamiltonian is always in the orthogonal class, and vice versa. We extract general principles regarding when a Floquet problem can thermalize to a different symmetry class to the instantaneous Hamiltonian. A (finite-sized) Floquet system can even display crossovers between different random matrix classes as a function of driving frequency. 
\end{abstract}
\maketitle

The quantum statistical mechanics of well isolated many body quantum systems is drawing intense interest, driven in part by recent experimental advances in the construction, control and measurement of such systems \cite{ref1, ref2}. One key question involves whether - and how - such well isolated quantum systems can thermalize. The Eigenstate Thermalization Hypothesis (ETH) \cite{Deutsch, Srednicki, RigolOlshanii} plays a central role in these discussions. For systems with eigenstates that obey the ETH, every (ETH obeying) many body eigenstate is individually in thermal equilibrium, in the sense that for a macroscopic system prepared in that eigenstate, the reduced density matrix for a small subregion equals the thermal density matrix, at a temperature set by the energy density in the eigenstate. These ideas have also been applied to periodically driven `Floquet' systems \cite{Floquet1, Floquet2}, which lack a conserved energy. In the absence of any conserved quantities, there arises a version of the ETH in which the reduced density matrix for a subregion is proportional to the {\it unit} matrix i.e. thermalization to `infinite temperatures.' Such periodically driven `Floquet' systems provide a particularly clean playground for investigations of quantum thermalization, and have inspired much recent work \cite{DAlessioRigol, Floquet4, Floquet5, Floquet6, Floquet7, ARCMP}. One question that has not been asked, however, is whether all thermalizing Floquet systems are the same, or if there exist sharply distinct infinite temperature phases. 

Random matrix theory \cite{Mehta} provides an independent and complementary approach to understanding thermalization. For systems that do thermalize, random matrix theory predicts that quantities such as eigenvalue statistics and level correlation functions should be governed by the relevant random-matrix ensemble - either one of the three traditional Wigner-Dyson ensembles, or, in the presence of particle-hole symmetry, the generalized Altland-Zirnbauer ensembles \cite{AltlandZirnbauer}.  Two Floquet systems described by distinct random matrix ensembles are in sharply distinct phases, even if both are at `infinite temperature.' However, the role of symmetries in determining the relevant random matrix ensemble for a Floquet system has not been explored.

In this Letter we explore the role of symmetries in Floquet thermalization. We ask: when a Floquet system thermalizes, can we determine the relevant random matrix ensemble by examining the symmetries of the (time dependent) Hamiltonian? We establish by analytical arguments and numerical exact diagonalization that the answer to the above question is {\it no}. The Floquet problem can thermalize, displaying random matrix statistics, even when the instantaneous Hamiltonian is always trivially integrable. A thermalizing Floquet system can also display an emergent integrability at certain commensurate driving frequencies. Even when the Floquet problem and the instantaneous Hamiltonian both thermalize, they can be in different thermalizing phases. In particular, the Floquet problem can be governed by the (circular) unitary ensemble even when the instantaneous Hamiltonian is governed by the (Gaussian) orthogonal ensemble at all times, and vice versa. We discuss under what situations the Floquet problem and the instantaneous Hamiltonian can thermalize to different symmetry classes. The Floquet problem can also display crossovers between othogonal and unitary regimes as a function of driving frequency (in addition to the well known crossovers between thermalizing and localized regimes). 

We restrict our discussion to the orthogonal and unitary ensembles, leaving extensions to the symplectic and Altland-Zirnbauer classes to future work.  Our results are obtained by working with `bang-bang' models, where the Hamiltonian is toggled between two discrete forms, since these provide the simplest realization of a Floquet system. However, we believe the conclusions to be generic. Our work is focused on level statistics as diagnostics of the random matrix class. We note also that unlike most work in the field, our calculations are not restricted to states in the middle of the spectrum - we are able to include states near the band edge by correcting for the varying density of states, using a normalization procedure introduced in \cite{Canali} (see Supplement for details). 

We focus on a simple model based on a chain of $N$ spins-$1/2$ with periodic boundary conditions. The instantaneous Hamiltonian is the generic anisotropic Heisenberg Hamiltonian with a random  field
\begin{eqnarray}
H\left(\{J_{\alpha}\},\{h_{\alpha}\}\right)&=& \sum_{\alpha=x,y,z} \left[ J_{\alpha} \sum_{i=1}^{N} S^{\alpha}_i S^{\alpha}_{i + 1}\right] \label{InstantHamiltonian}\\
&&+\sum_{\alpha=x,y,z}  \left[h_{\alpha} \sum_{i=1}^{N} c_{\alpha,i} S^{\alpha}_i \right] \nonumber
\end{eqnarray}
where $S^{\alpha}_i = \frac12 \sigma^{\alpha}_i$ and where $\sigma^{\alpha}$ is a Pauli matrix. The coeffients $c_{\alpha,i}$ are uncorrelated and chosen according to a uniform distribution within the interval $[-1,1]$. The amplitude of the random  field is set through the $h_{\alpha}$. This model exhibits various level statistics depending on the parameters $\{J_{\alpha}\}$ and $\{h_{\alpha}\}$. Setting all parameters to zero except $J_z$ and $h_z$ leads to a trivially integrable model. If we now take $J_x=J_y=J_z=1$ and $h_x=h_y=0$, $h_z = h$ then we obtain the spin half Heisenberg model with random $z$ fields, which is a workhorse of studies of many body localization. This model displays a many body localized phase (with Poisson level statistics) for large $h \gtrsim 3.5$, and a thermalizing phase (with GOE level statistics in a sector with fixed total $S^z$) for small $h$ \cite{Pal}. Note that the level statistics are GOE even though the time reversal symmetry is broken by the  field because of the presence of a disguised antiunitary symmetry, made up of time reversal and a rotation by $\pi$ of all spins about the $x$ axis, which leaves the Hamiltonian unchanged. Similarly, if we allow two components of the field to be non-zero (e.g. $h_x \neq 0$, $h_y \neq 0$, $h_z = 0$), then too the  level statistics are described by the Gaussian orthogonal ensemble (GOE), for small fields when the system thermalizes. The relevant antiunitary symmetry is now time reversal plus a $\pi$ rotation about the $z$ axis (i.e. $S^x \rightarrow S^x$, $S^y \rightarrow S^y$, $S^z \rightarrow - S^z$), which leaves the Hamiltonian unchanged \cite{Avishai}. Once all three fields are non-zero however there is no longer any such antiunitary symmetry, and the level statistics in the thermalizing phase are described by the Gaussian unitary ensemble (GUE) (see Supplement).

The Hamiltonian of Eq.~\ref{InstantHamiltonian} is the building block of our `bang-bang' model. We focus on the two bang case, which is the simplest possible structure for a Floquet problem. The time dependent $\tau$-periodic Hamiltonian ${\cal H}_{\rm 2 bangs}(t)$ is defined  as
\begin{eqnarray}
{\cal H}_{\rm 2 bangs}\left(0 < t < \frac{\tau}{2}\right)& = H_1 =&H\left(\{J_{\alpha,1}\},\{h_{\alpha,1}\}\right),\nonumber\\
{\cal H}_{\rm 2 bangs}\left(\frac{\tau}{2} < t < \tau\right)& = H_2 =&H\left(\{J_{\alpha,2}\},\{h_{\alpha,2}\}\right).\label{TwoBangHamiltonian}
\end{eqnarray}
where $\{J_{\alpha,1}\}$ and $\{h_{\alpha,1}\}$ (resp.  $\{J_{\alpha,2}\}$ and $\{h_{\alpha,2}\}$) set the ${\cal H}_{\rm 2 bangs}(t)$ when $0< (t\;{\rm mod}\; \tau) < \frac{\tau}{2}$ (resp. $\frac{\tau}{2}< (t\; {\rm mod}\; \tau) < \tau$). Note that the random coefficients $c_{\alpha,i}$ are \emph{identical} for $H_1$ and $H_2$. The time evolution operator corresponding to Hamiltonian evolution over $\tau$ is 
\begin{equation}
U(\tau) = \exp(-i H_1 \tau/2) \exp(-i H_2 \tau/2).\label{TwoBangEvolution}
\end{equation}
Our goal is to explore the connection between the statistics of the eigenvalues $e^{i \lambda_n}$ of $U(\tau)$ and the level statistics of the time dependent instantaneous Hamiltonian ${\cal H}$. 

To numerically probe the level statistics, we compute the ratio of adjacent gaps. For a sorted spectrum $\{\lambda_n;\lambda_n \leq \lambda_{n+1}\}$, the ratio of adjacent gaps is defined as
\begin{eqnarray}
r_n&=&\frac{{\rm min}\left(\lambda_n - \lambda_{n-1}, \lambda_{n+1} - \lambda_{n}\right)}{{\rm max}\left(\lambda_n - \lambda_{n-1}, \lambda_{n+1} - \lambda_{n}\right)}\label{RatioAdjacentGap}
\end{eqnarray}
The definition above is strictly valid only when the density of states is constant, which which requires projecting onto the middle of the spectrum \cite{Pal}. We use an alternative `normalized' definition for $r $ which corrects for an energy dependent density of states (see Supplement for details), which allows us to retain all the states in our calculation of $ r $. Depending on the level statistics, the average ratio $r$ of adjacent gaps is $r \simeq 0.530$ for COE\cite{DAlessioRigol}, $r \simeq 0.60$ for CUE \cite{DAlessioRigol} and $r \simeq 0.386$ for a Poisson spectrum \cite{Pal}. Since we are restricted to moderate matrix sizes, we also average $r$ over different samples, denoting by $\langle r \rangle$ the corresponding ensemble averaged value.

As a warm-up, let's consider the situation where both $H_1$ and $H_2$ are integrable by using $J_{x,1}=J_{z,2}=1$, $h_{x,1}=h_{z,2}=h$ and all the other parameters being zero, i.e.
\begin{eqnarray}
H_1&=&\sum_{i=1}^{N} S^{x}_i S^{x}_{i + 1}+ h c_{x,i} S^{x}_i,\label{H1Integrable}\\
H_2&=&\sum_{i=1}^{N} S^{z}_i S^{z}_{i + 1}+ h c_{z,i} S^{z}_i.\label{H2Integrable}
\end{eqnarray}

\begin{figure}[htb]
\includegraphics[width=0.7\linewidth]{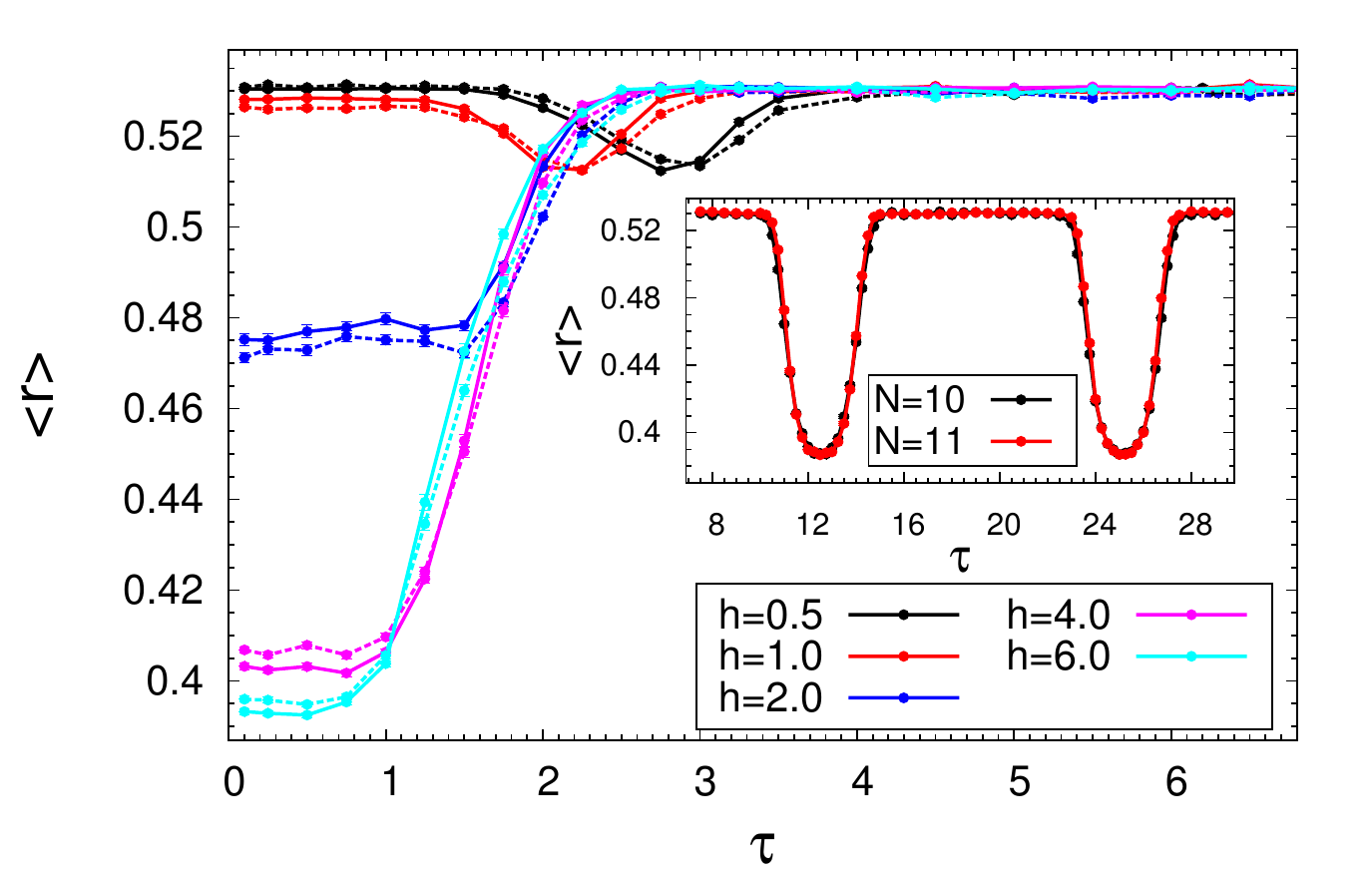}
\caption{Average ratio of adjacent gaps for the Floquet unitary of the two bang-models where $H_1$ and $H_2$ are given by Eqs.~\ref{H1Integrable} and \ref{H2Integrable}. The calculations were done for $N=10$ spins and various amplitudes of the random field $h$. $\langle r \rangle$ has been averaged over 200 samples. Inset: Results for the same model at larger $\tau$, showing an emergent integrability at commensurate frequencies $\tau = 4 n \pi$. While we only show the data for $h=0.5$, the results are identical at least up to $h=6.0$.}
\label{IntegrableCOE}
\end{figure}

As can be observed in Fig.~\ref{IntegrableCOE}, the Floquet problem thermalizes to the orthogonal ensemble for any value of $\tau$, as long as the random fields $h$ are weak enough. Thermalization is governed by the orthogonal ensemble because the Floquet operator (\ref{TwoBangEvolution}) is invariant under the antiunitary symmetry $S^x \rightarrow S^x$, $S^y \rightarrow S^y$, $S^z \rightarrow - S^z$ as discussed above. Note that the Floquet problem can thermalize even though the instantaneous Hamiltonian is always integrable, because the constants of motion of the instantaneous Hamiltonian change over time, such that the Floquet Hamiltonian does {\it not} have an extensive number of local constants of motion. 

For large fields the model displays a many body localized phase in the high frequency limit, diagnosed by Poisson level statistics. However, the system always thermalizes below a (field dependent) critical frequency, which is the generic behavior for Floquet systems with an MBL phase \cite{Lazarides} and is related to the increasingly non-local response to time dependent local perturbations \cite{nonlocal}. 

 Note the existence of a dip in $\langle r \rangle $ at a field strength dependent value of $\tau \approx 2$. The origin of this dip (and similar dips in Figs. 2,3,4) is discussed at length in the supplementary material (and in Ref.\cite{DAlessioRigol}). In brief, the dip occurs when the Floquet zone width first becomes comparable to the many body bandwidth (such that states start getting `folded' into the principal Floquet zone) and seems to be a universal signature of the weakened level repulsion between states that have and have not been reconstructed by the resulting many body resonances. 

Finally, note that the model discussed above actually displays {\it integrable} behavior at a discrete set of frequencies $\tau= 4 n \pi$ (integer $n$). This emergent integrability is discussed at length in the Supplementary Material, and illustrates the special behavior that can arise in Floquet problems at commensurate frequencies. 

\begin{figure}[htb]
\includegraphics[width=0.7\linewidth]{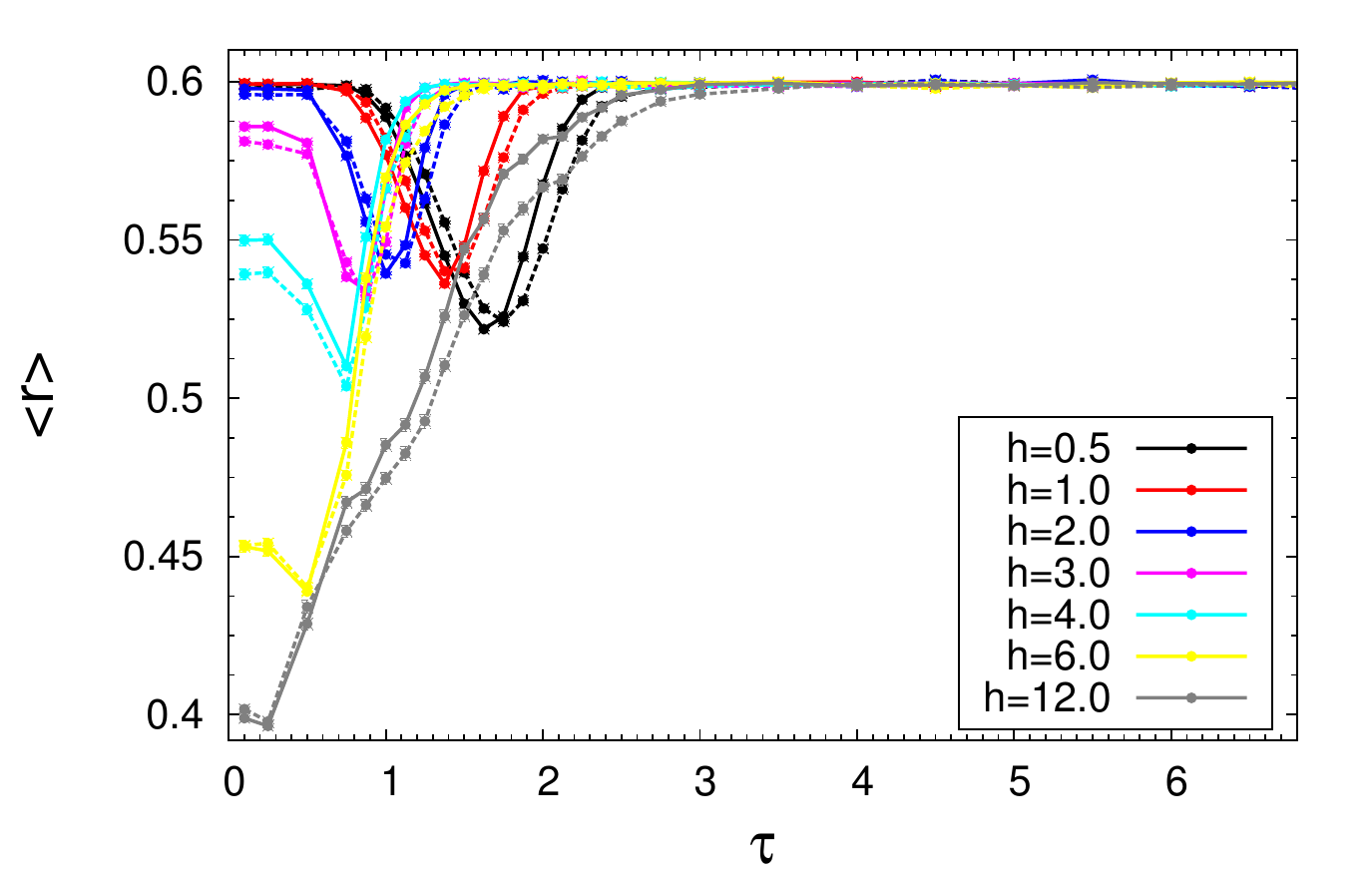}
\caption{$\langle r \rangle$ value for the Floquet unitary operator of the two bang-models where $H_1$ and $H_2$ are GOE and given by Eqs.~\ref{H1GOE} and \ref{H2GOE}. The calculations were done for $N=11$ (solid lines) and $N=10$ (dashed lines) spins and various amplitudes of the random  field $h$. $\langle r \rangle$ has been averaged over 300 samples.}
\label{GOEFullCUE}
\end{figure}

We now discuss situations where the instantaneous Hamiltonian and the Floquet problem are both thermalizing, and discuss the (lack of) any relation between the relevant symmetry classes for thermalization. We begin by pointing out that the Floquet problem can thermalize to the CUE for all $\tau$ even when the instantaneous Hamiltonian always thermalizes to the orthogonal class. This can be achieved e.g. in a model with 
\begin{eqnarray}
H_1&=&\sum_{i=1}^{N} \vec{S}_i \cdot \vec{S}_{i + 1}+ h \left(c_{x,i} S^{x}_i + \frac{1}{2} c_{y,i} S^{y}_i\right),\label{H1GOE}\\
H_2&=&\sum_{i=1}^{N} \vec{S}_i \cdot \vec{S}_{i + 1}+ h \left(\frac{1}{2} c_{y,i} S^{y}_i + c_{z,i} S^{z}_i\right).\label{H2GOE}
\end{eqnarray}
The instantaneous Hamiltonian only ever has a  field along two axes, is thus invariant under an appropriate antiunitary transformation, and thus at weak disorder thermalizes to the orthogonal ensemble $\langle r \rangle \approx 0.53$ (see Supplementary Material). In contrast, the Floquet problem involves  fields along all three axes, and is not invariant under any such antiunitary transformation, and thus thermalizes to the unitary ensemble (Fig.~\ref{GOEFullCUE}), at least for weak fields. For stronger  fields there exists a localized phase with Poisson statistics for high driving frequencies, which gives way to a thermalizing phase in the CUE class for low frequencies (see Supplementary Material for details). We can understand thermalization of the Floquet problem to the unitary class as follows: in the model discussed above, for $H_1$ the relevant antiunitary symmetry is the improper rotation $S^z \rightarrow - S^z$, whereas for $H_2$ it is $S^x \rightarrow - S^x$. However, since $H_1$ and $H_2$ have different antiunitary symmetries, there is no antiunitary symmetry for $U(\tau)$. We believe this result - that the Floquet Hamiltonian can be CUE even if the instantaneous Hamiltonian is GOE if the antiunitary symmetries change over time - is general, and not particular to two bang models.  

We can also have a situation where the instantaneous Hamiltonian is always GUE but the Floquet problem is COE. In a two bang model with equal length bangs, this happens if there is an antiunitary symmetry that exchanges $H_1$ and $H_2$. In this case the antiunitary symmetry leaves $U(\tau)$ unchanged, up to a shift of $\tau/2$ in the origin of time, even while it changes the instantaneous Hamiltonian. A specific example is a model with all the $J_{\alpha} = 1$ and  $h_{\alpha,1}=h_{x,2}=h_{z,2}=-h_{y,2}=h$, wherein the two Hamiltonians $H_1$ and $H_2$ are transformed into one another by the improper rotation $S^y \rightarrow - S^y$. This can be seen to have COE level statistics (Fig.\ref{GUECOECUE}) even through the instantaneous Hamiltonians are GUE ({\bf $\langle r \rangle = 0.6$} - see Supplementary material). Again we believe this result to be general - even if the instantaneous Hamiltonian is not invariant under any antiunitary symmetry, if the Floquet operator is so invariant (up to a shift in the origin of time) then the Floquet level statistics will be COE.  This may be useful numerically, since driven systems with GUE instantaneous Hamiltonians may nevertheless be represented by a completely real Floquet Hamiltonian $H_F = \frac{-i}{\tau} \ln U(\tau)$. Of course, this emergent antiunitary symmetry of the stroboscopic time evolution operator can be broken by applying $H_1$ and $H_2$ for unequal times $\tau(1/2 \pm \epsilon)$, in which case the Floquet problem reverts to the unitary class (Fig.~\ref{GUECOECUE}).

\begin{figure}
\includegraphics[width=0.7\linewidth]{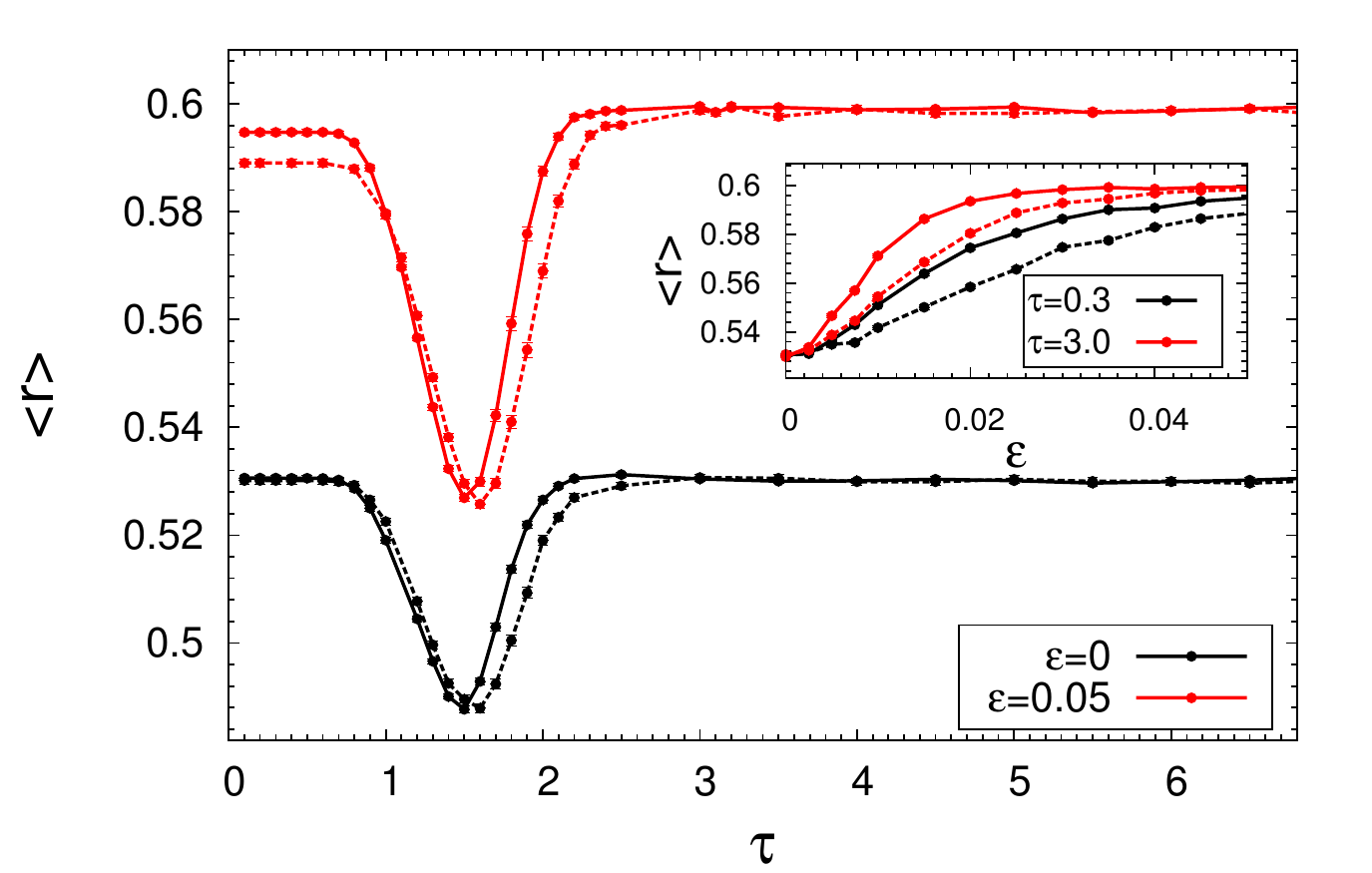}
\caption{The Floquet Hamiltonian of the two bang-models where $H_1$ and $H_2$ transform into one another under an antiunitary transformation,  where the time spent evolving with $H_1$ is $\tau (1/2 - \epsilon)$ and the evolution with $H_2$ is performed during  $ \tau (1/2 + \epsilon)$. We consider both system sizes $N=11$ spins (solid lines) and $N=10$ (dashed lines) spins. The black lines corresponds to the case $\epsilon=0$ while the red lines are for $\epsilon =0.05$. The inset show the evolution of $\langle r \rangle$ when changing $\epsilon$ for $\tau=0.3$ (black lines) and  $\tau=3.0$ (red lines).}\label{GUECOECUE}
\end{figure}

\begin{figure}[htb]
\includegraphics[width=0.7\linewidth]{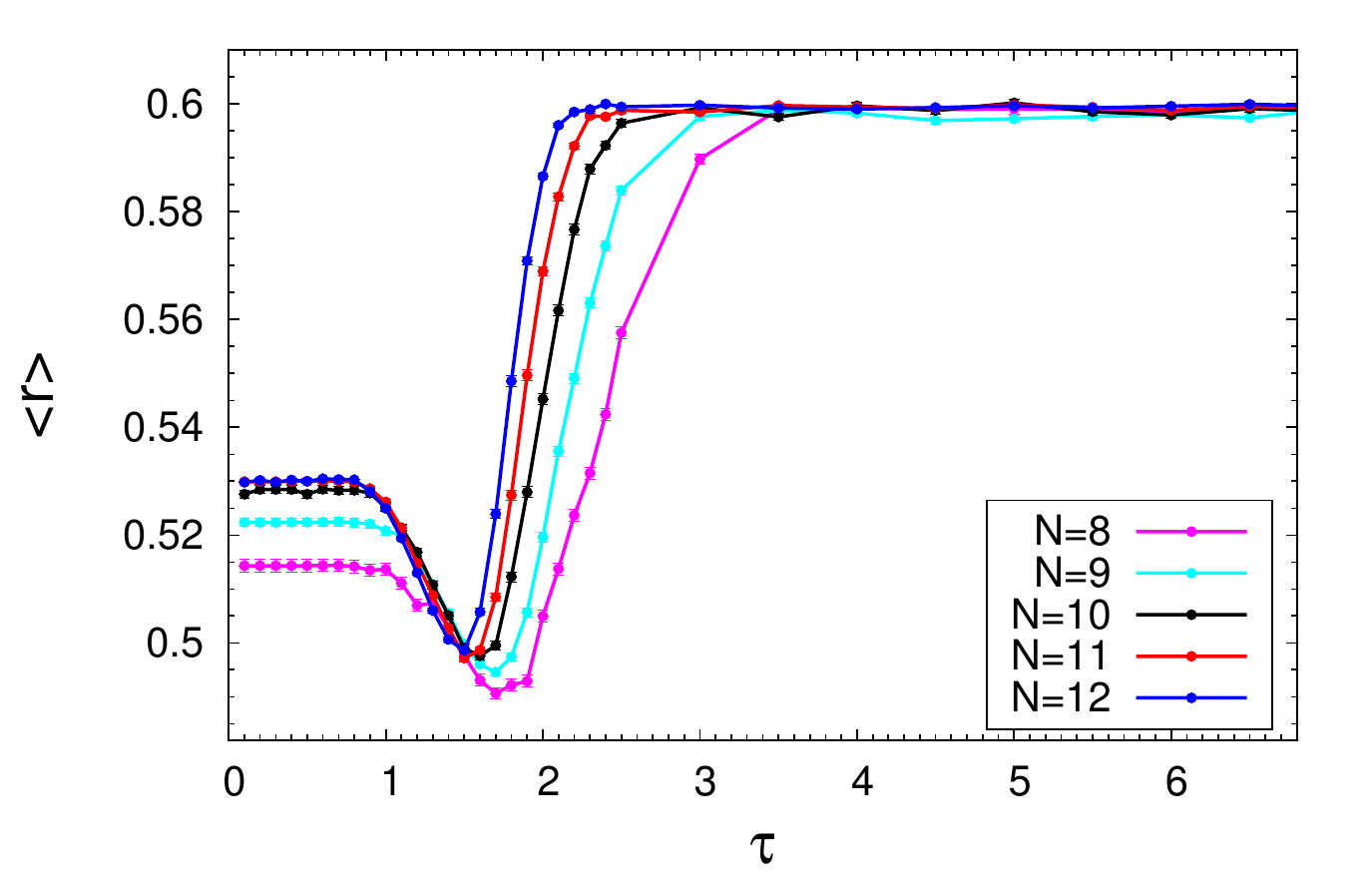}
\caption{Average ratio of adjacent gaps for the Floquet Hamiltonian of the two bang-models (Eq.2) with all $J=1$ and $(h_{x,1}, h_{y,1}, h_{z,1}) = (0.5,0.5,0)$ and $(h_{x,2}, h_{y,2}, h_{z,2}) = (0,-0.5,0.5)$, for systems of $N$ spins. $\langle r \rangle$ has been averaged over 200 samples.  }
\label{GOECOECUE}
\end{figure}

 We now discuss situations where the Floquet Hamiltonian displays transitions between random matrix classes as a function of driving frequency. One model that does this has all $J=1$ and $(h_{x,1}, h_{y,1}, h_{z,1}) = (0.5,0.5,0)$ and $(h_{x,2}, h_{y,2}, h_{z,2}) = (0,-0.5,0.5)$. This has fields along all three axes so in general should be in the unitary class, but should have orthogonal statistics in the $\tau \rightarrow 0$ limit since $H_1 + H_2$ has vanishing field along the $y$ axis. What we see numerically (Fig.~\ref{GOECOECUE}) is however much more striking. There is a finite regime of frequencies $\tau< \tau_c \approx 1$ over which we observe orthogonal statistics, with unitary statistics not setting in until $\tau \gtrsim 2 \tau_c$. If the dip in $\langle r \rangle$ is identified with the onset of folding states into the principal Floquet zone, then the `orthogonal regime' is presumably the regime when the bandwidth is less than the Floquet zone width. The frequency window over which this is true should shrink to zero in the thermodynamic limit, since the bandwidth of an interacting system is an extensive quantity. However, the shrinking of the size of this window with system size is extremely slow (Fig.4), and thus an appreciable `orthogonal regime' may be seen in modest sized systems. It is interesting to note that resonances between Floquet states in different zones (in an extended zone scheme) are apparently essential to drive this $U(\tau)$ from the orthogonal to the unitary class. 
 
{\it Conclusions:} Our numerical investigation of Floquet thermalization in the orthogonal and unitary symmetry classes reveals the following general principles: (i) the Floquet problem can thermalize, displaying random matrix statistics, even when the instantaneous Hamiltonian is always integrable, if the instantaneous constants of motion change over time. In this case the Floquet problem can display anomalous integrable behavior when the driving frequency is commensurate with characteristic energy scales in the instantaneous integrable Hamiltonians.(ii)  A thermalizing instantaneous Hamiltonian $H(t)$ will be governed by the orthogonal ensemble IFF it is invariant under an anti-unitary symmetry transformation $\mathcal{T}(t)$. However, even if the instantaneous Hamiltonian is always governed by the orthogonal ensemble, the Floquet problem can be in the unitary class IFF the instantaneous antiunitary $\mathcal{T}(t)$ changes as a function of time. (iii) The Floquet problem can be in the orthogonal class even if the instantenous Hamiltonian is in the unitary class, if there exists an antiunitary transformation which leaves the Floquet operator unchanged up to a shift in the origin of time. (In a two bang model this happens if the antiunitary transformation exchanges $H_1$ and $H_2$). (iv) There can arise crossovers between orthogonal and unitary thermalization as a function of driving frequency. Our results apply to the entire spectrum, not just the states in the middle of the band. Extensions to other symmetry classes and continuously time varying Hamiltonians are left to future work. 

{\it Acknowledgments:} We acknowledge fruitful discussions with D. Huse. N.R. was supported by the Princeton Global Scholarship.

\newpage
\begin{widetext}
\section{Supplementary Material to ``Floquet Thermalization: Symmetries and Random Matrix Ensembles''}

In this Supplementary Material, we provide additional analytical and numerical results that might be relevant to a more specialized audience. The Supplementary Material is broken up into sections addressing particular issues that were mentioned in the main text. 

\subsection{Normalization procedure for retaining full spectrum in level statistics calculations}
Most works in the field concentrate on states near the middle of the many body spectrum. This is necessary because the density of states varies significantly over the full spectrum (as can be seen for example in Fig.~\ref{SupplMat:DOS}a) , and the changing density of states effect spoils conventional measures of level statistics. Restricting to an energy window near the middle of the spectrum (where the density of states is nearly constant) ameliorates this problem. In this work, however, we make use of an alternative normalization prescription apparently first introduced by \cite{Canali}. This allows us to use the full spectrum for calculating level statistics, without needing to project out any states. We note that \cite{Canali} referred to this normalization procedure as `unfolding' but we do not use this term here. 

Starting from the original energies $\lambda$ with an average density of states $\langle\rho(\lambda)\rangle$, we introduce a new variable $s(\lambda)$ such that its average density is constant i.e. $\langle\tilde{\rho}(s)\rangle=1$. This variable is defined by the relation
\begin{equation}
s(\lambda)=\int_{-\infty}^{\lambda}d\lambda' \; \langle\rho(\lambda')\rangle\label{SupplMat:UnfoldingVariable}
\end{equation}
The calculation of the average ratio of adjacent gaps is performed as following. We first compute the average density of states $\langle\rho(\lambda)\rangle$ of the Floquet Hamiltonian over the different samples. Then for each sample, we transform the Floquet Hamiltonian spectrum from the $\lambda$ variable to the $s$ variable. The ratios of adjacent gaps are computed for each sample in the new variable $s$
\begin{eqnarray}
r_n&=&\frac{{\rm min}\left(s_n - s_{n-1}, s_{n+1} - s_{n}\right)}{{\rm max}\left(s_n - s_{n-1}, s_{n+1} - s_{n}\right)}\label{SupplMat:RatioAdjacentGapUnfoldingVariable}
\end{eqnarray}
from which we deduce the average ratio $r$ of adjacent gaps. $r$ is then averaged over the different samples to obtain $\langle r \rangle$. All $r$ values reported in the main text are calculated from the full spectrum using this normalization procedure. 

\subsection{Localization-delocalization transitions for Hamiltonians in the unitary symmetry class}

Consider the Hamiltonian
\begin{equation}
H\left(\{J_{\alpha}\},\{h_{\alpha}\}\right)=\sum_{\alpha=x,y,z} \left[ J_{\alpha} \sum_{i=1}^{N} S^{\alpha}_i S^{\alpha}_{i + 1}\right] \label{SupplMat:InstantHamiltonian} +\sum_{\alpha=x,y,z}  \left[h_{\alpha} \sum_{i=1}^{N} c_{\alpha,i} S^{\alpha}_i \right].
\end{equation}
where we take $J_x=J_x=J_z=1$. The coeffients $c_{\alpha,i}$ are uncorrelated and chosen according to a uniform distribution within the interval $[-1,1]$. The amplitude of the random  field is set through the $h_{\alpha}$. When all the $h_{\alpha}$ are equal to zero, the hamiltonian is integrable. If the random field is restricted to the same plane irrespective of the site, the hamiltonian exhibit an anti-unitary symmetry (a combination of the time-reversal symmetry and a $\pi$ rotation about the spin axis orthogonal to the plane). As a consequence for moderate  field, the level statistics satisfy the GOE. The transition from GOE to the Poisson statistics when increasing the  field has been extensively studied especially in the case where $h_x=h_y=0$ to preserve $S^t_z$ the total spin along the $z$-axis. Such a symmetry allows to consider larger spin chains in numerical simulations. The transition occurs around $h\simeq 3.5 \pm 1.0$ \cite{Pal}. For our purpose, we need to consider cases where the $S^t_z$ is not conserved. The absence of $S^t_z$ as a conserved quantity also restricts the system sizes that can be studied (here up to $N=12$). For completeness, we have thus studied the case where $h_x=h_z=h$ and $h_y=0$, looking at the evolution of the ratio of adjacent gaps $\langle r \rangle$ as a function of $h$. The calculations were performed up to $N=12$ sites using periodic boundary conditions and 400 samples. As can be seen in Fig.~\ref{SupplMat:GOEGUE}a, we observe a transition from GOE to Poisson statistics around $h\simeq 3$ (as judged by when the level statistics parameter gets halfway between its Poisson and GOE values) , which corresponds to a total random field strength of $\simeq 3 \sqrt{2} = 4.2$ once we account for the fact that we now have fields along two axes. This is within the estimated range for the model with fields along the $z$ axis only i.e. we do not observe any noticeable difference in the critical field strength for the localization-delocalization transition when we have fields along two axes rather than only one. This problem is of course still in the orthogonal class, as has been discussed in the main text. 
\begin{figure}[htb]
(a)\includegraphics[width=0.45\linewidth]{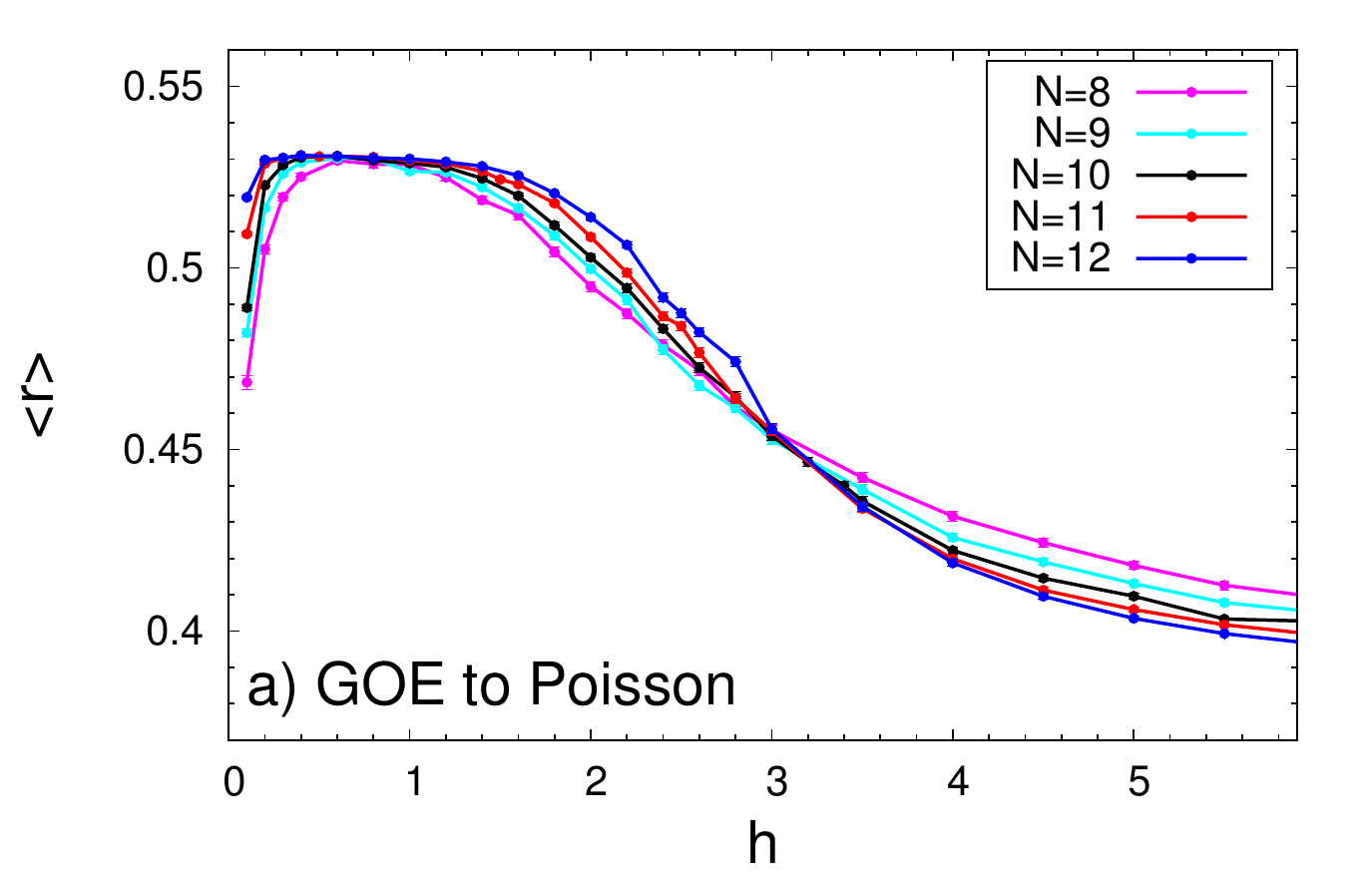}
(b) \includegraphics[width=0.45\linewidth]{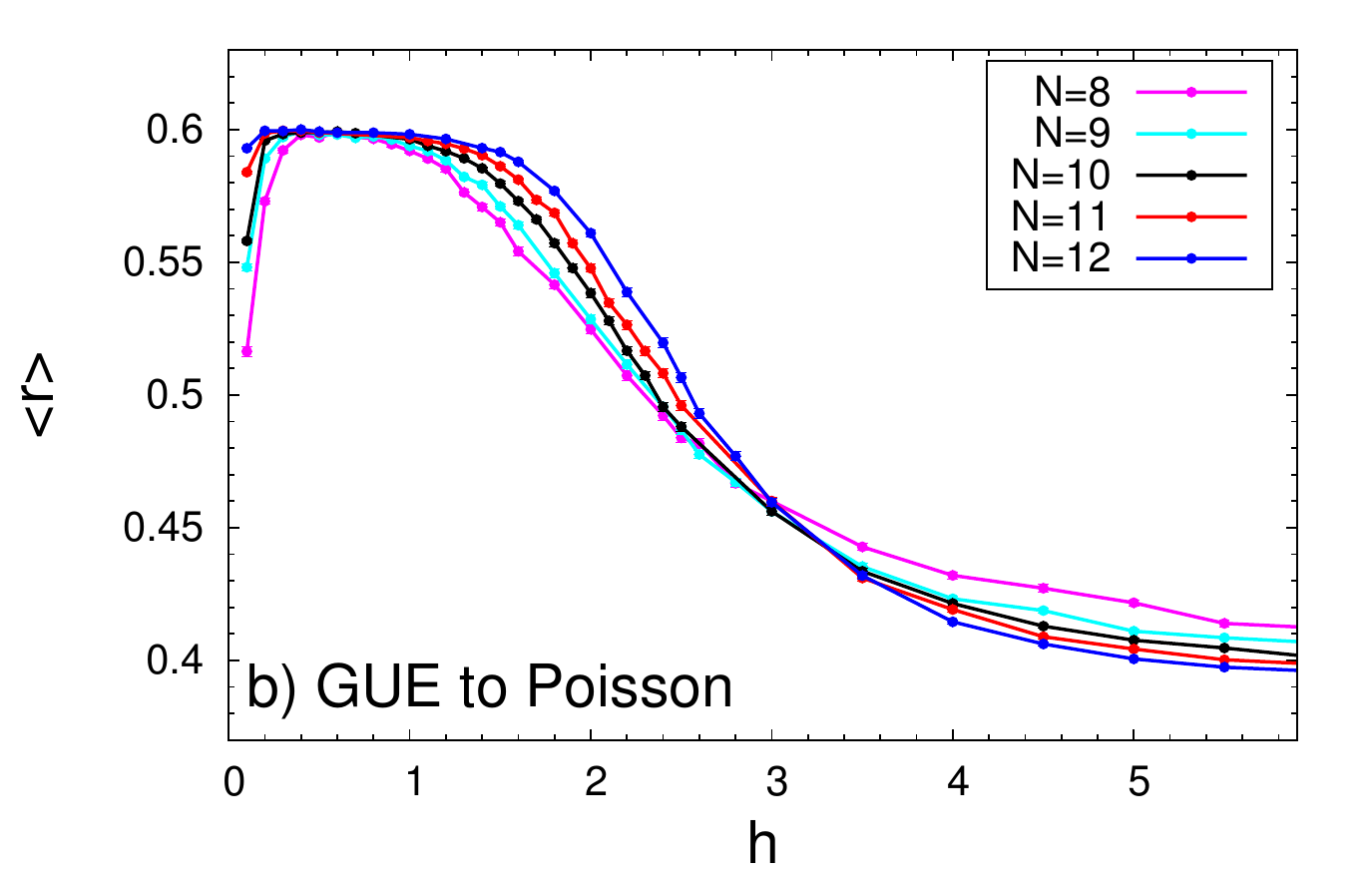}
\caption{(a): Transition from GOE to Poisson statistics as a function of the amplitude of the random  field $h$ along the $x$ and $z$ axes. (b) Localization delocalization transition in the unitary class: transition from GUE to Poisson statistics as a function of the amplitude of the random  field $h$ along the $x$, $y$ and $z$ axes. Data obtained on $N=10, 11$ and $12$ spin chains with periodic boundary conditions. }
\label{SupplMat:GOEGUE}
\end{figure}

We also investigated the (Hamiltonian) localization delocalization transition in the unitary class, when the random  field can be oriented along any direction in three dimensional space (i.e. not restricted to a plane). As shown in Fig.~\ref{SupplMat:GOEGUE}b there is once again a transition from random matrix statistics (this time GUE) to Poisson statistics driven by the strength of the random field (effectively by disorder strength). This time the level statistics parameter gets halfway between its GUE and Poisson values around $h \simeq 2.5$, corresponding to a total field strength of $2.5 * \sqrt{3} = 4.3$ i.e. we do not observe any sharp difference between the critical disorder strength for localization in the unitary and orthogonal classes. 

\subsection{Emergent integrability at commensurate frequencies}
Even when a Floquet problem thermalizes for general driving frequencies, it can nonetheless display integrable behavior for a discrete set of `commensurate' driving frequencies. We illustrate this with a discussion of a model of the form discussed in the main text: a `two bang' model with 
\begin{eqnarray}
U(\tau) &=& \exp(-i H_1 \tau/2) \exp(-i H_2 \tau/2) \nonumber\\
H_1&=&\sum_{i=1}^{N} S^{x}_i S^{x}_{i + 1}+ h c_{x,i} S^{x}_i,\\
H_2&=&\sum_{i=1}^{N} S^{z}_i S^{z}_{i + 1}+ h c_{z,i} S^{z}_i
\end{eqnarray}
and with periodic boundary conditions. Some elementary algebraic manipulations allow us to rewrite the Floquet operator as 
\begin{eqnarray}
U(\tau) &=& \exp\big(-i \frac{\tau}{2} \sum_{j=1}^{N} (S^{x}_j S^{x}_{j+ 1}+ h c_{x,j} S^{x}_j)\big)\exp\big(-i \frac{\tau}{2} \sum_{k=1}^{N} (S^{z}_k S^{z}_{k+ 1}+ h c_{z,k} S^{z}_k)\big)\nonumber\\
&=& \exp\big(-i \frac{\tau}{2} \sum_{j=1}^{N} S^{x}_j S^{x}_{j+ 1} \big)\exp \big(-i h \frac{\tau}{2} \sum_k c_{x,k} S^{x}_k\big)\exp\big(-i \frac{\tau}{2} \sum_{l=1}^{N} S^{z}_l S^{z}_{l+ 1}\big)\exp \big(-i h \frac{\tau}{2} \sum_m c_{z,m} S^{z}_m\big) \nonumber\\
&=& \prod_{j=1}^{N}\exp\big(-i \frac{\tau}{2} S^{x}_j S^{x}_{j+ 1} \big)  \prod_{k=1}^{N}\exp \big(-i h \frac{\tau}{2} c_{x,k} S^{x}_k\big)  \prod_{l=1}^{N}\exp\big(-i \frac{\tau}{2}  S^{z}_l S^{z}_{l+ 1}\big)  \prod_{m=1}^{N}\exp \big(-i h \frac{\tau}{2} c_{z,m} S^{z}_m\big) \nonumber\\
&=& \prod_{j=1}^{N}\exp\big(-i \frac{\tau}{8} \sigma^{x}_j \sigma^{x}_{j+ 1} \big)  \prod_{k=1}^{N}\exp \big(-i h \frac{\tau}{4} c_{x,k} \sigma^{x}_k\big)  \prod_{l=1}^{N}\exp\big(-i \frac{\tau}{8}  \sigma^{z}_l \sigma^{z}_{l+ 1}\big)  \prod_{m=1}^{N}\exp \big(-i h \frac{\tau}{4} c_{z,m} \sigma^{z}_m\big) \nonumber\\
&=& \prod_{j=1}^{N} \big( \cos \frac{\tau}{8} - i \sin \frac{\tau}{8} \sigma^x_j \sigma^x_{j+1} \big)  \prod_{k=1}^{N}\big( \cos \frac{h \tau c_{x,k}}{4} - i \sin \frac{h \tau c_{x, k}}{4} \sigma^x_k \big) \nonumber \\
&\times& \prod_{l=1}^{N} \big( \cos \frac{\tau}{8} - i \sin \frac{\tau}{8} \sigma^z_l \sigma^z_{l+1} \big)  \prod_{m=1}^{N}\big( \cos \frac{h \tau c_{z,m}}{4} - i \sin \frac{h \tau c_{z, m}}{4} \sigma^z_m \big) \end{eqnarray}
where we recall that the $S^{\alpha}$ are spin half operators $S^{\alpha} = \frac12 \sigma^{\alpha}$. Now one can straightforwardly see that driving frequencies when $\tau =  4 n \pi$ (integer $n$) are special, since in this case the stroboscopic time evolution operator becomes simply 
\begin{equation}
U(\tau) = \exp \big(-i h \frac{\tau}{2} \sum_k c_{x,k} S^{x}_k\big)\exp \big(-i h \frac{\tau}{2} \sum_m c_{z,m} S^{z}_m\big).
\end{equation}
In this stroboscopic time evolution operator, every spin is decoupled, and the stroboscopic time evolution simply performs a (different and independent) rotation on every spin. There are thus an extensive number of local conserved quantities (one per spin, associated with the projection of that spin along the effective rotation axis) and an emergent integrability - but only for commensurate driving frequencies $\tau = 4 n \pi$. Away from these commensurate frequencies, the problem does thermalize, as illustrated by Fig.1 of the main text.

\subsection{Orthogonal-Unitary crossovers}
Here we discuss models that should display at least a crossover between orthogonal and unitary regimes driven by frequency. An example of a model that does this is one with all $J=1$, with $(h_{x,1}, h_{y,1}, h_{z,1}) = (h,h,0)$ and with $(h_{x,2}, h_{y,2}, h_{z,2}) = (0, -h, h)$. For general $\tau$ this corresponds to a Floquet Heisenberg model with fields along all three axes, which should be in the unitary class. However, for $\tau \rightarrow 0$ the Floquet Hamiltonian is just $H_1+H_2$, which has fields only along $x$ and $z$ axes (the fields along the $y$ axis in $H_1$ and $H_2$ are equal and opposite and thus cancel out). We thus expect a crossover to orthogonal statistics as we take the limit $\tau \rightarrow 0$. What we see in numerics - shown in Fig.~\ref{SupplMat:GOECOECUE} - is however much more striking. There is a finite window of $\tau \lesssim \tau_c \approx 1$ wherein we have orthogonal statistics for weak fields, with unitary statistics only setting in above $\tau \gtrsim 2 \tau_c$. It is possible that $\tau_c$ will shrink to zero in the thermodynamic limit \cite{DAlessioRigol}, but even if this does happen the shrinking of $\tau_c$ with $N$ is exceedingly slow, and a noticeable regime of orthogonal statistics may well be visible at high frequencies for modest sized systems. 

\begin{figure}[htb]
\includegraphics[width=0.3\linewidth]{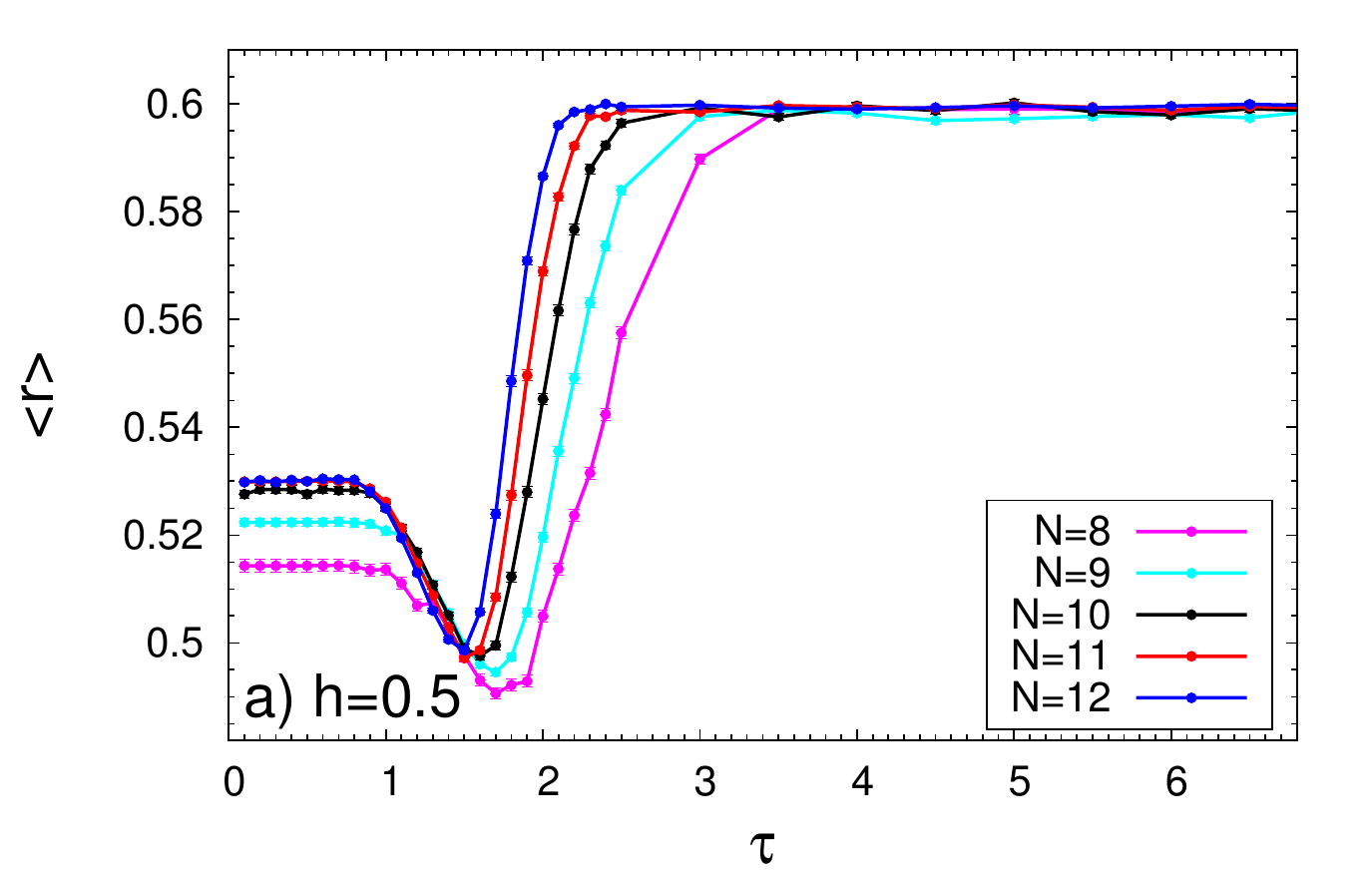}
\includegraphics[width=0.3\linewidth]{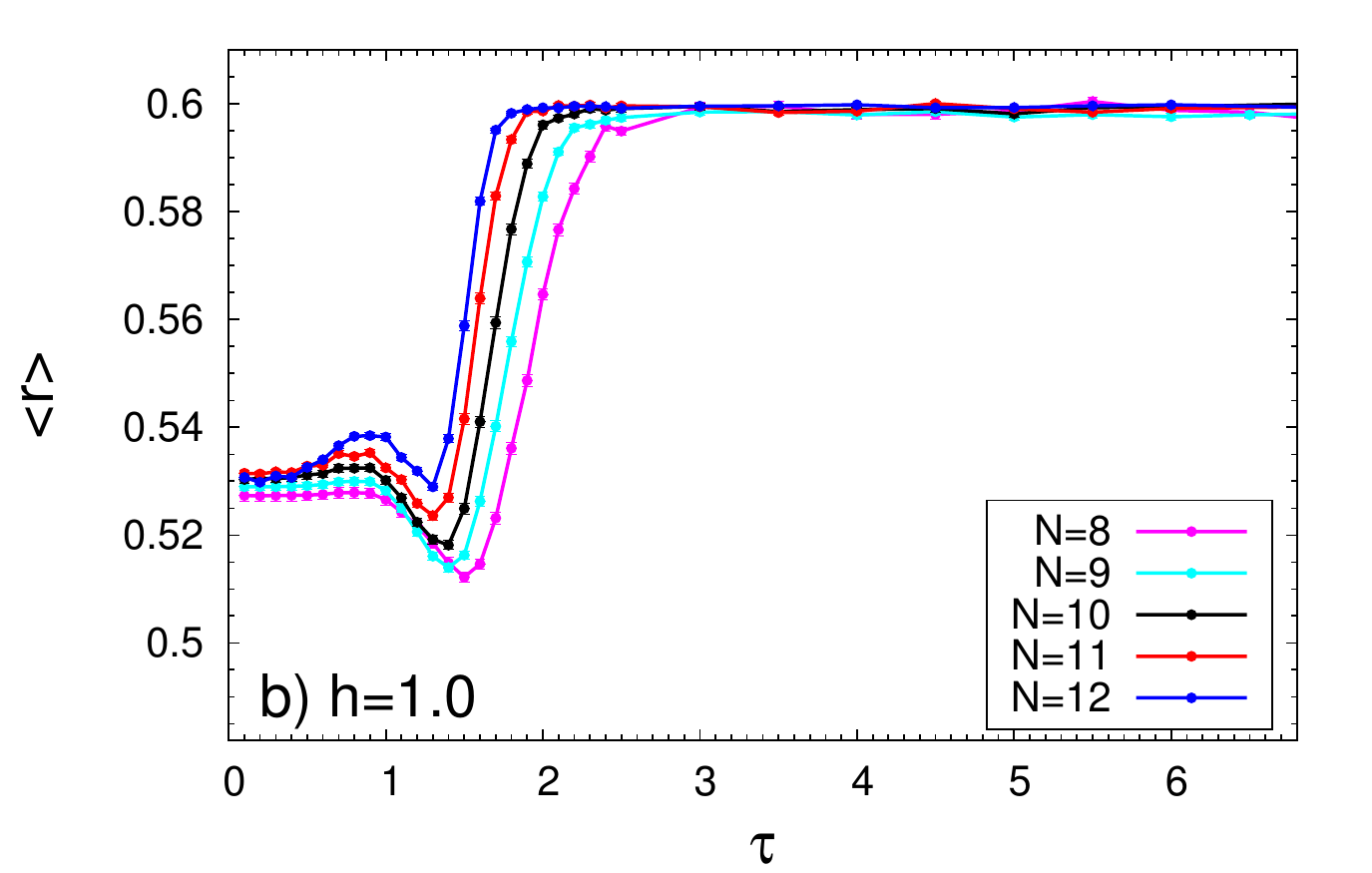}
\includegraphics[width=0.3\linewidth]{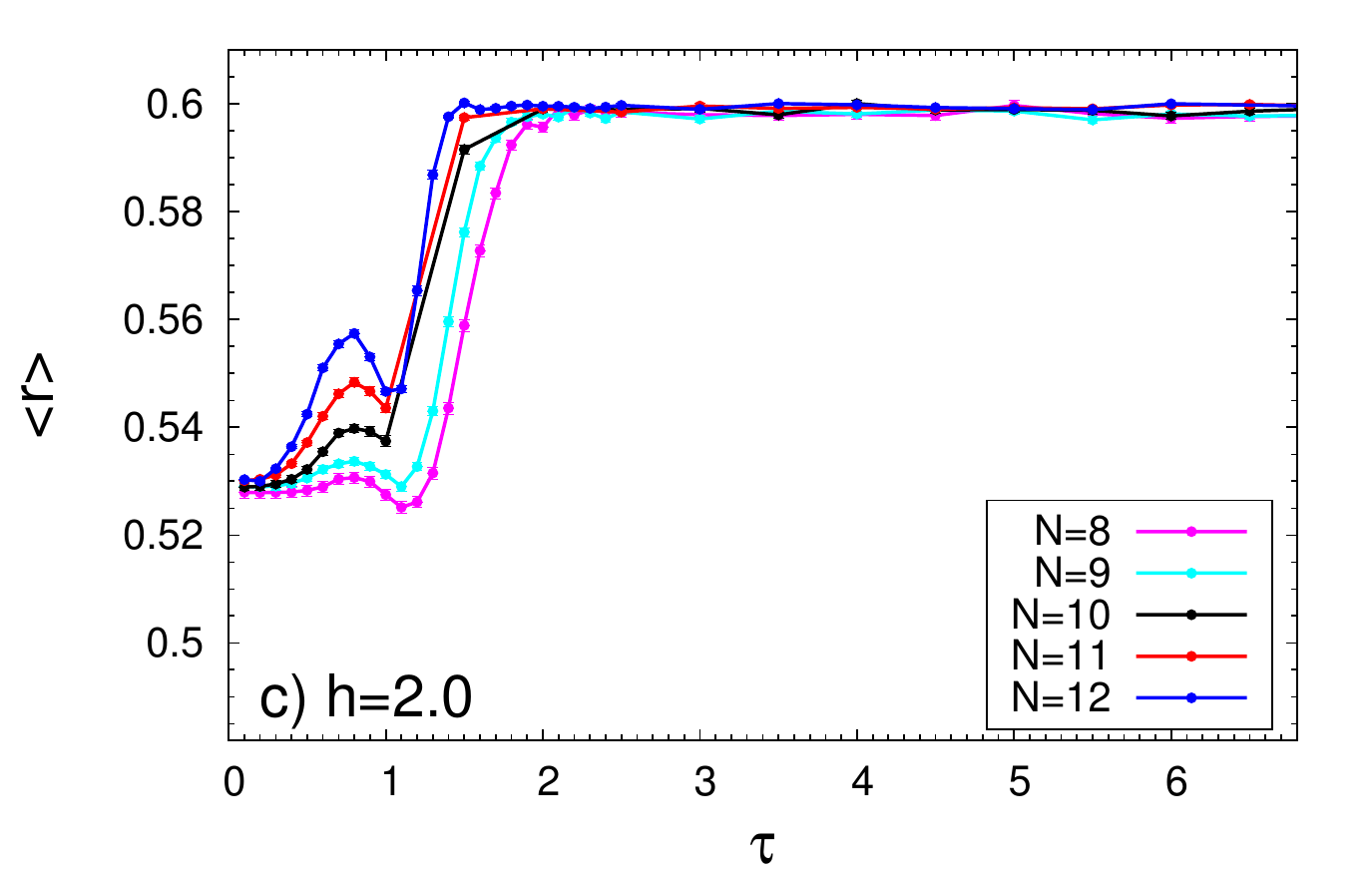}
\includegraphics[width=0.3\linewidth]{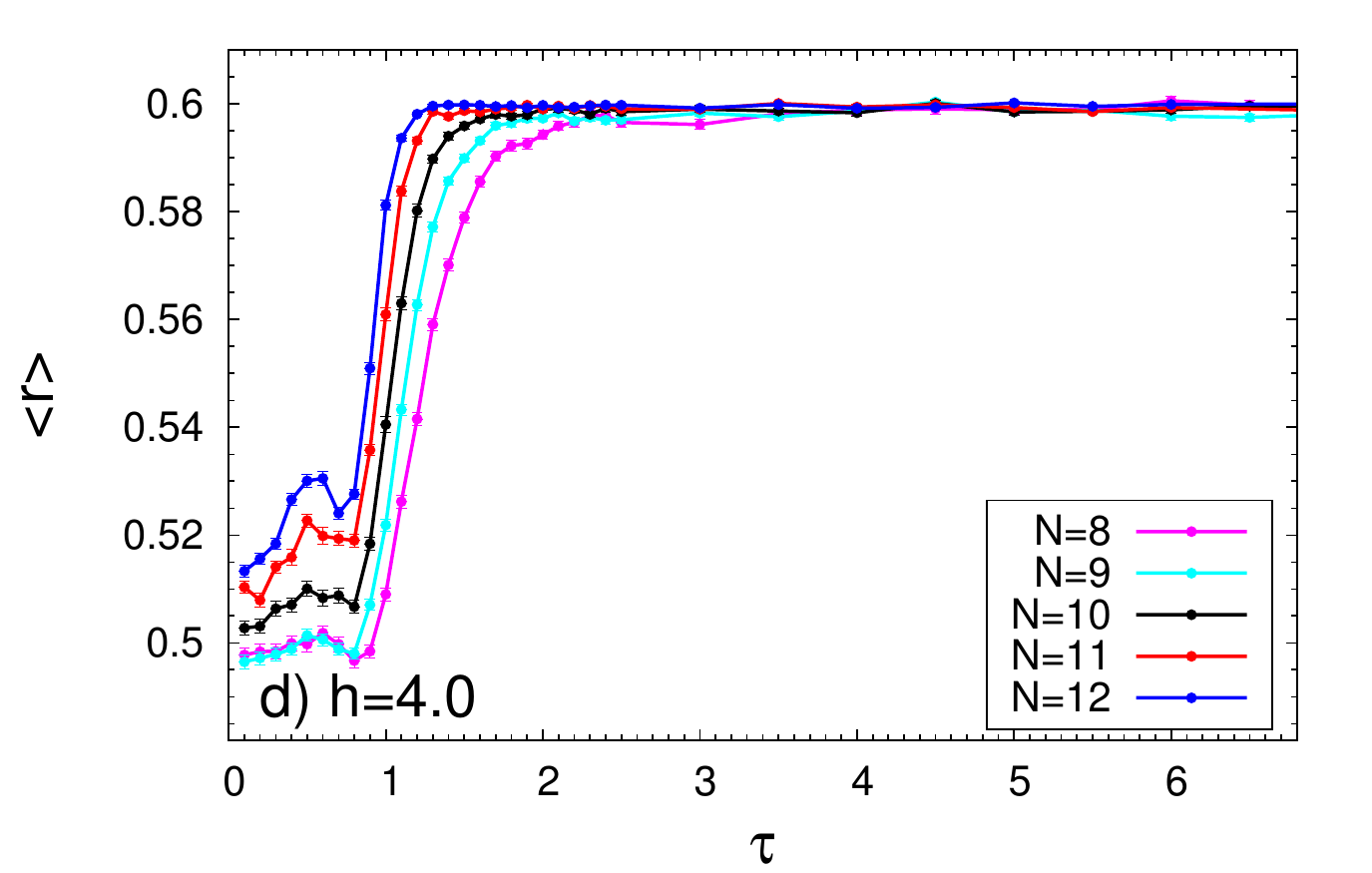}
\includegraphics[width=0.3\linewidth]{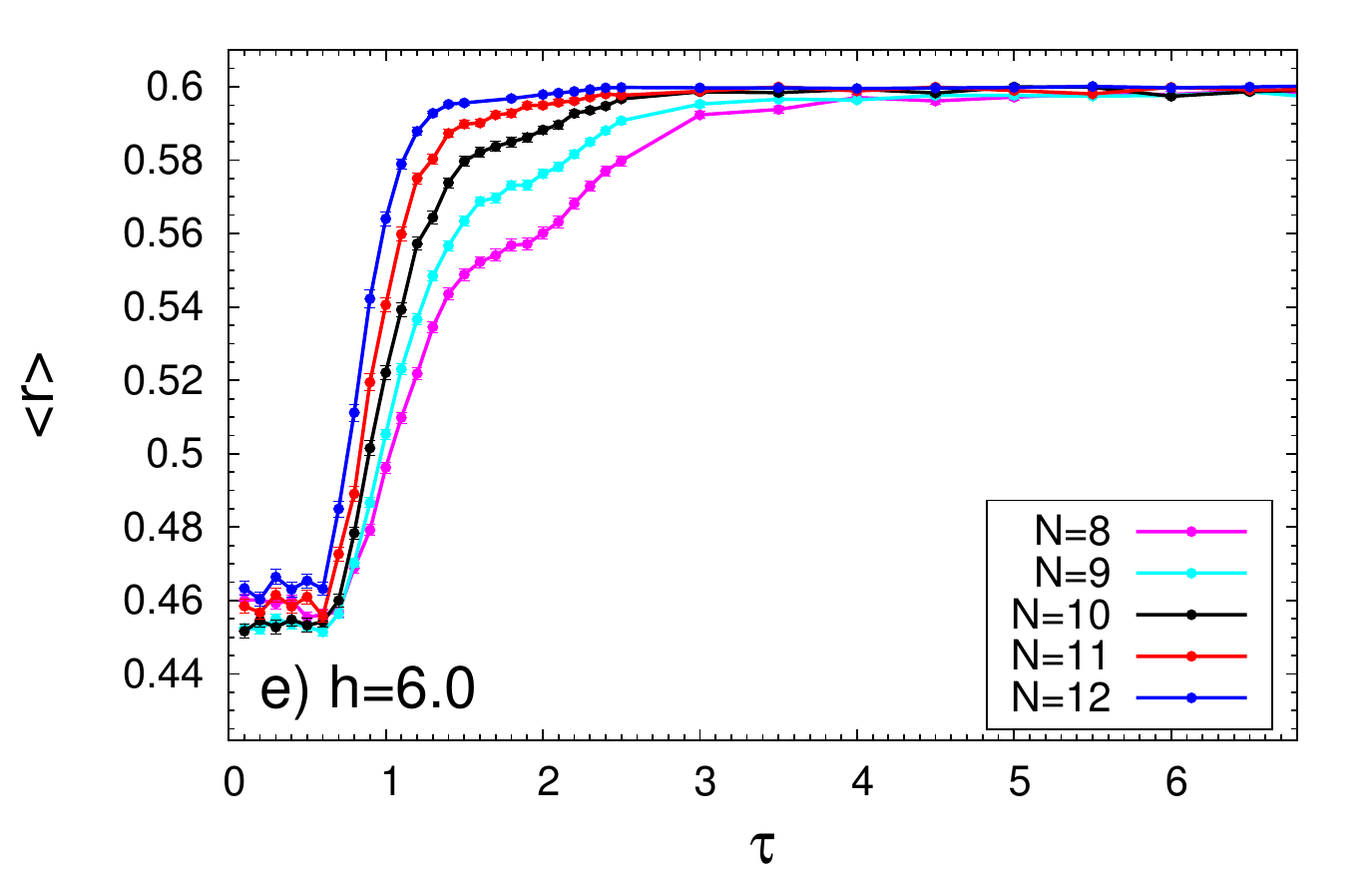}
\includegraphics[width=0.3\linewidth]{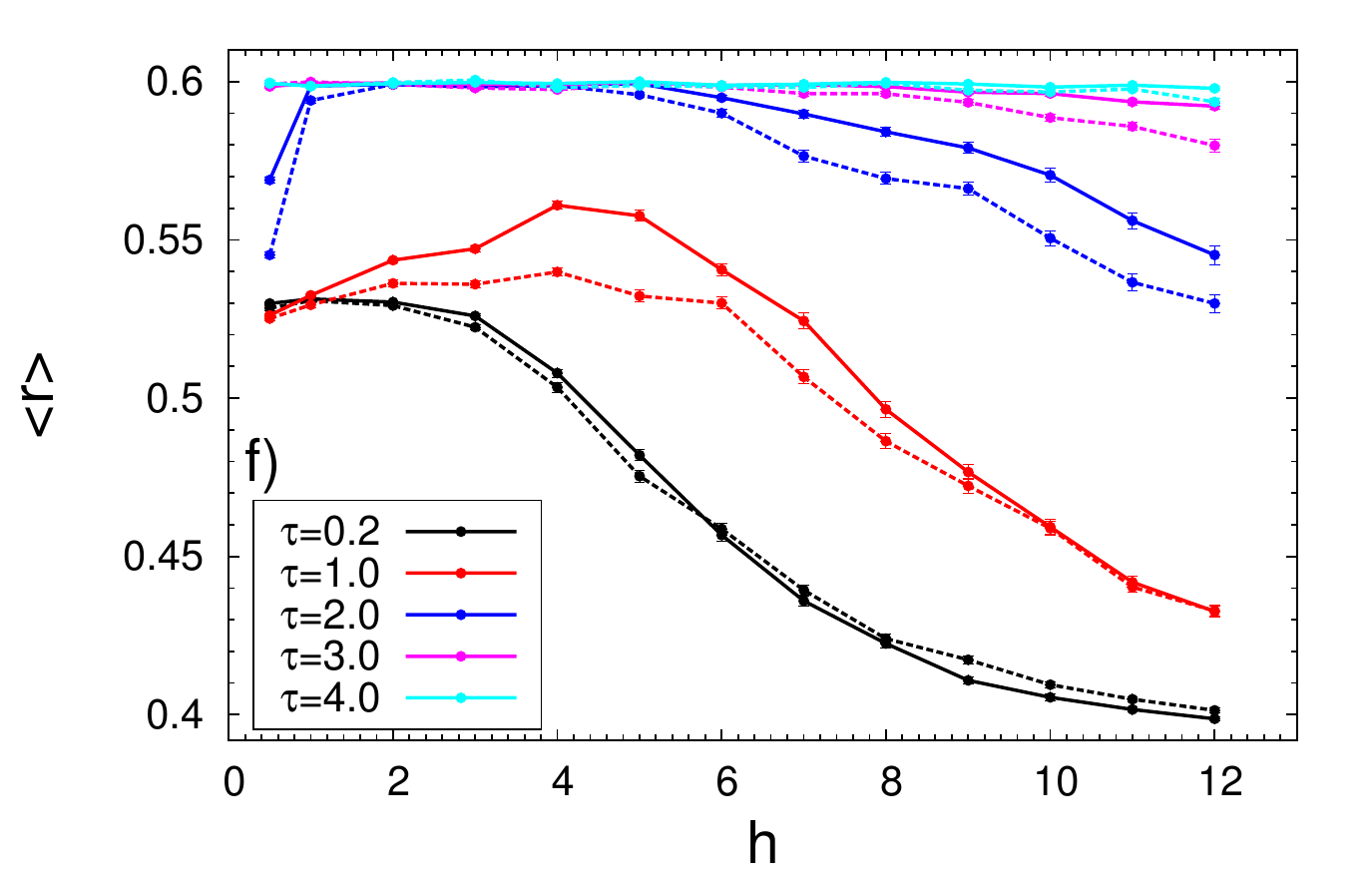}
\caption{Transition from COE to CUE in the two bang model from $N=8$ up to $N=12$ spins and various amplitudes of the random  field. From top to bottom : $h=0.5$ (a), $h=1.0$ (b), $h=2.0$ (c), $h=4.0$ (d) and $h=6.0$ (e). We have used 200 samples for $N=10,11$ and $12$ and 400 samples for $N=8$ and 9. The large $\tau$ limit (at fixed field strength) always thermalizes to the CUE. For strong disorder, the small $\tau$ limit starts to show onset of localization (d, e), but for weak disorder the small $\tau$ regime is COE (a-c). We stress (a) that at weakest disorder, the COE to CUE crossover is quite sharp. Finally, (f) shows the evolution of the level statistics parameter in the two bang model for $N=11$ (plain lines) and $N=10$ (dashed lines) spins as a function of the disorder strength $h$ for various driving periods $\tau$. Note that the weak disorder limit is COE for small $\tau$ but CUE for large $\tau$.
\label{SupplMat:GOECOECUE}}
\end{figure}

\begin{figure}[htb]
\includegraphics[width=0.45\linewidth]{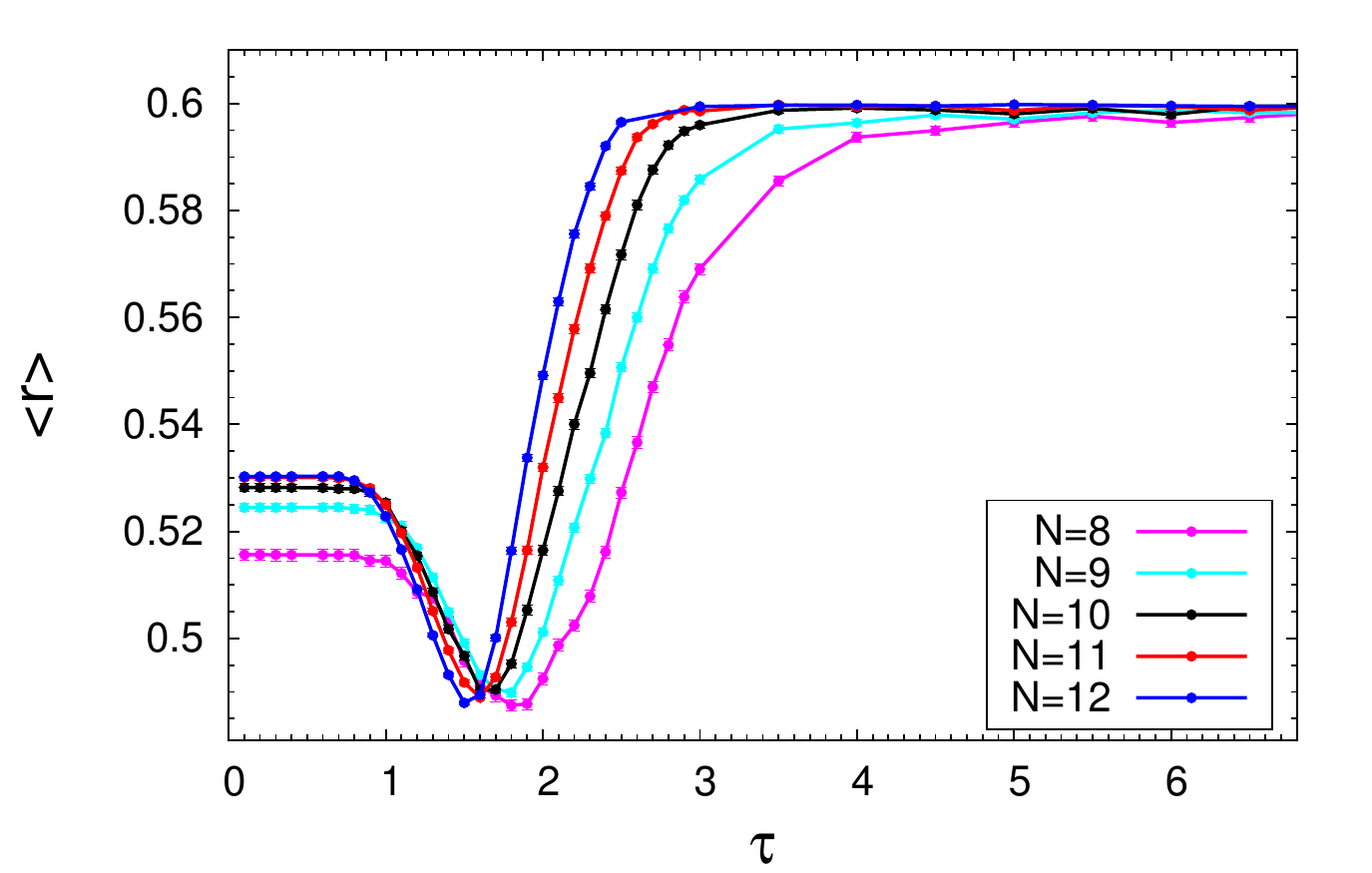} 
\caption{Transition from COE to CUE statistics controlled by driving frequency in a two bang model where $H_1$ and $H_2$ almost transform into one another under an antiunitary transformation (but not quite), and where $H_1 + H_2$ is in the orthogonal class.}\label{SupplMat:GUECOECUE}
\end{figure}

Another possibility is to consider a two bang model where $H_1$ and $H_2$ are both GUE but (almost) transform into one another under an antiunitary transformation, up to a slight mismatch in field strengths. For example, we could consider a model with all $J = 1$ and $h_{x,1}=0.5+ \epsilon$, $h_{x,2}=0.5 - \epsilon$, $h_{z,1}=0.5 - 2 \epsilon$, $h_{z,2}=0.5+ 2 \epsilon$, and $h_{y,1}=-h_{y,2}=1$. For $\epsilon = 0$, this model is in the orthogonal class for all $\tau$ since $H_1$ and $H_2$ are interchanged under improper spin rotation. For {\it any} $\epsilon$, COE statistics should be recovered in the $\tau \rightarrow 0$ limit, when the Floquet Hamiltonian is just $H_1 + H_2$ (which has  fields restricted to the $x-z$ plane). For $\epsilon \neq 0$ and $\tau \neq 0$ we would expect in general to observe CUE statistics, with a crossover to orthogonal behavior in the $\tau \rightarrow 0$ limit. Numerically, for $\epsilon = 0.1$ we do observe COE behavior at small $\tau$ and CUE at large $\tau$ (Fig.\ref{SupplMat:GUECOECUE}), but the behavior does {\it not} look like a crossover to the $\tau \rightarrow 0$ limit. Rather we appear to see a finite regime of stability of orthogonal statistics, with a fairly sharp crossover to unitary statistics around $\tau_c \approx 1$. Again the dependence of $\tau_c$ on system size is fairly weak, and an appreciable orthogonal regime may be visible at high frequencies for modest system sizes. 

\subsection{Universal dips in $\langle r \rangle$ when the Floquet zone width becomes of order the bandwidth. }
\begin{figure}[htb]
\includegraphics[width=0.45\linewidth]{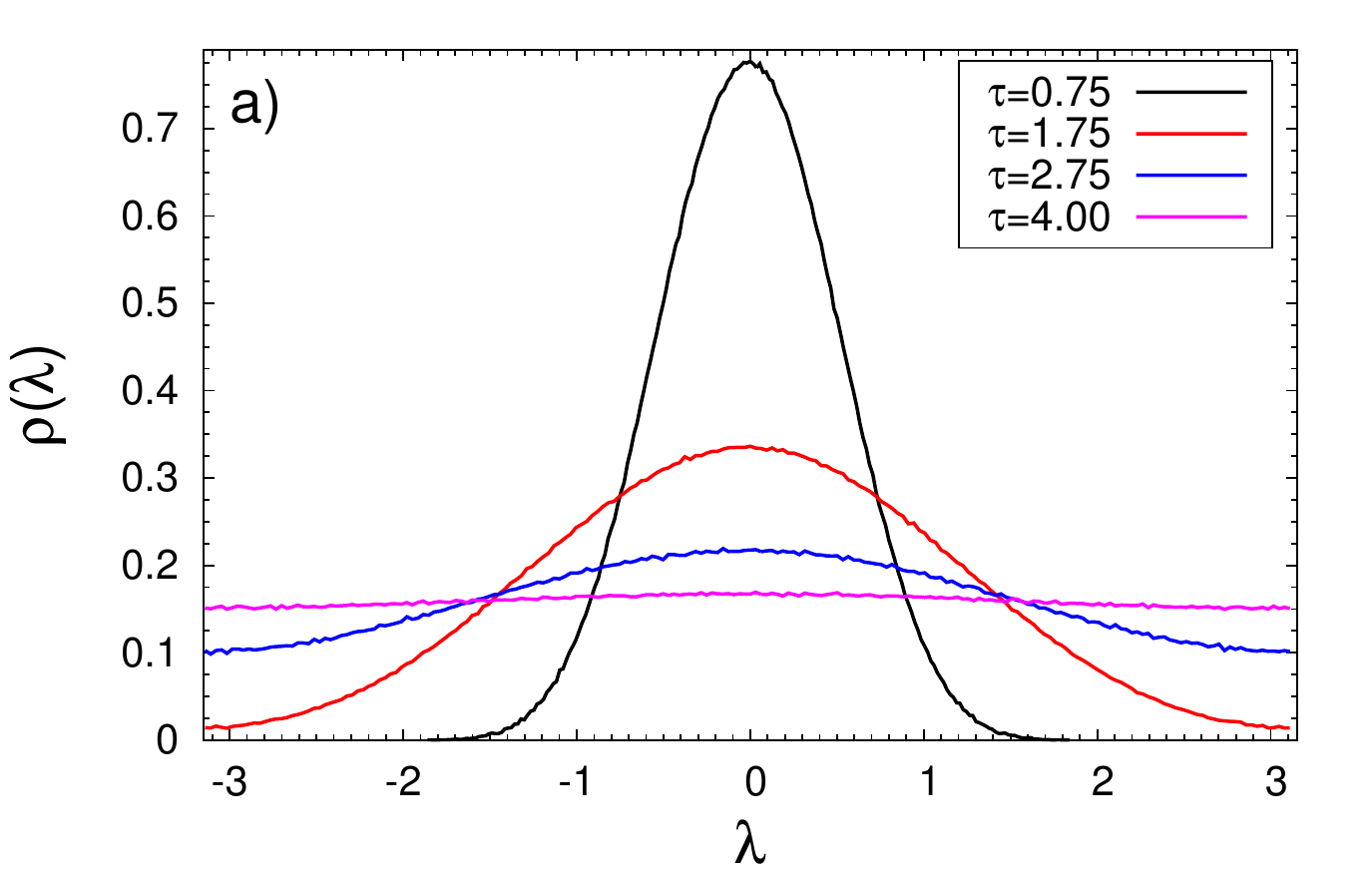} 
\includegraphics[width=0.45\linewidth]{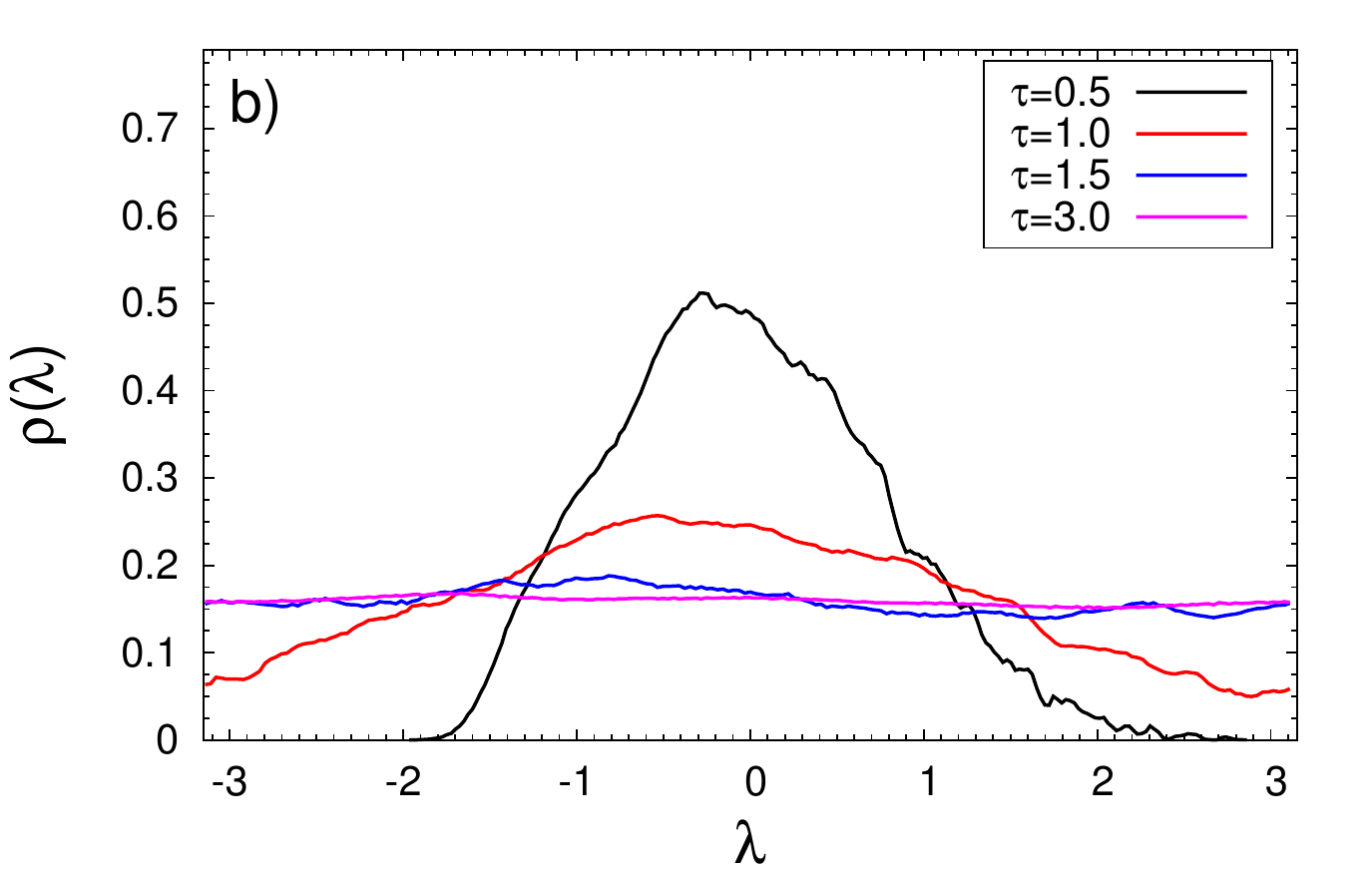} 
\caption{Density of eigenphases $\lambda$ of $U(\tau)$ averaged over 200 samples at various period $\tau$. (a) the two bang-models where $H_1$ and $H_2$ are integrable and given by Eqs.~\ref{H1Integrable} and \ref{H2Integrable} for $N=11$ spins with a random  field amplitude $h=0.5$. (b) the two bang-models where $H_1$ and $H_2$ are GOE but $H_1+H_2$ is GUE, for $N=12$ spins with a random  field amplitude $h=0.5$.}\label{SupplMat:DOS}
\end{figure}
One universal feature of all the Floquet models studied is a dip in the $\langle r \rangle$ value at a particular driving frequency. To understand this, we look at how the density of Floquet eigenphases changes as a function of driving frequency (Fig. \ref{SupplMat:DOS}). At high frequency, the density of eigenphases is contained entirely within the principal Floquet zone $[-\pi,\pi]$. We can thus `unfold' to an extended zone scheme with no ambiguity - the bandwidth is less than the Floquet zone width, and all the eigenphases lie within the principal zone. This is the regime in which high frequency expansions converge. it was argued \cite{DAlessioRigol} that in this the Floquet Hamiltonian is well approximated by the time averaged Hamiltonian. There is full level repulsion between all eigenstates, as long as the time averaged Hamiltonian is thermalizing. The width of this zone in $\tau$ should shrink to zero in the thermodynamic limit, since the bandwidth of a thermalizing system is an extensive quantity. 

In contrast, at low frequency (large $\tau$), the bandwidth is much greater than the Floquet zone width (the density of states is essentially uniform across the principal Floquet zone) and we cannot unambiguously unfold to an extended zone scheme. This is the regime in which the Floquet Hamiltonian is very different from the time averaged Hamiltonian, and all the eigenstates have been extensively reconstructed (i.e. look very different to the eigenstates of the time averaged Hamiltonian \cite{DAlessioRigol}). Again there is full level repulsion between all eigenstates. In the thermodynamic limit, this will presumably be the only regime. 

The dip in $\langle r \rangle$ starts when the Floquet zone width first becomes of order the bandwidth (i.e. when the eigenphases first start to wrap around the principal zone). This happens e.g. when $\tau \approx 1.75$ for the model in Fig.1 and Fig.8(a), and when $\tau \approx 1$ for the model in Fig.2 and Fig.8b. The dip ends and the low frequency regime begins when the density of eigenphases becomes uniform across the principal Floquet zone i.e. when $\tau \approx 4$ for the model in Fig.1 and Fig.8a, and around $\tau \approx 3$ for the model in Fig.2 and Fig.8b. The dip thus happens precisely when the eigenenergies of the time averaged Hamiltonian are starting to get folded back into the principal zone, but when the folding back is not yet reached the middle of the principal zone. Another way to say this is that in quasienergy windows near the edges of the principal zone, states that would have lived in different zones in an extended zone scheme have been folded on top of each other, but in the center of the principal zone the only states are states that would have been there even in an extended zone scheme. 

Now the eigenstates of the time averaged Hamiltonian have level repulsion between then, but the bandwidth of the time averaged Hamiltonian is larger than the Floquet zone width, so the band edges of the time averaged Hamiltonian get folded back into the principal Floquet zone. After folding back, eigenstates of the time averaged Hamiltonian that were essentially uncorrelated (and had very different spatial structure) end up near degenerate in quasi-energy. The true eigenstates of $U(\tau)$ are thus massively reconstructed by resonances between eigenstates of the time averaged Hamiltonian with very different spatial structure, and look completely different to the eigenstates of the time averaged Hamiltonian \cite{DAlessioRigol}. Again, there is full level repulsion between reconstructed eigenstates (and thus $\langle r \rangle $ is governed by random matrix theory in the low frequency limit). 

However, in the intermediate frequency regime when the dip in $\langle r \rangle$ happens, {\it not} all the states have been reconstructed, since only a small fraction of the band of the time averaged Hamiltonian has been folded back into the principal Floquet zone. While there is full level repulsion between unreconstructed states and between reconstructed states, there is {\it not} full level repulsion between reconstructed and unreconstructed states, which both co-exist in this intermediate frequency regime. We believe that this lack of full level repulsion between reconstructed and unreconstructed states is responsible for the intermediate frequency dip in $\langle r \rangle$, with the minimum of the dip occurring when roughly half the states have been reconstructed. This dip seems to thus be a robust and highly universal finite size effect in thermalizing Floquet systems. 

\end{widetext}

\end{document}